\documentclass[aps,pre,twocolumn,showpacs]{revtex4-1}
\usepackage[utf8]{inputenc} 
\usepackage[english]{babel}
\usepackage{fullpage}
\usepackage{graphicx}
\usepackage{fancyvrb}
\usepackage{units}
\usepackage{xcolor}
\usepackage{lmodern}
\usepackage{amsmath,amsfonts}
\usepackage{mathtools}
\usepackage{array}
\usepackage{indentfirst}
\usepackage{perpage} 
\MakePerPage{footnote} 
\usepackage[babel=true]{csquotes} 
\usepackage{soul}

\DeclarePairedDelimiter\abs{\lvert}{\rvert}%
\begin{document}

\title{Nonlinear Optical Galton Board: thermalization and continuous limit}
\author{Giuseppe Di Molfetta}
\affiliation{LERMA, Observatoire de Paris, PSL Research University, CNRS, Sorbonne Universités, UPMC Univ. Paris 6, UMR 8112, F-75014, Paris France}
\author{Fabrice Debbasch}
\affiliation{LERMA, Observatoire de Paris, PSL Research University, CNRS, Sorbonne Universités, UPMC Univ. Paris 6, UMR 8112, F-75014, Paris France}
\author{Marc Brachet}
\affiliation{Laboratoire de Physique Statistique de l'Ecole Normale 
Sup{\'e}rieure  / PSL Research University, \\
associ{\'e} au CNRS et aux Universit{\'e}s Pierre-et-Marie-Curie Paris 06 et Paris Diderot,\\
24 Rue Lhomond, 75231 Paris, France}
\date{\today}
\begin{abstract}
The nonlinear optical Galton board (NLOGB), a quantum walk like (but nonlinear) discrete time quantum automaton, is shown to admit a complex evolution leading to long time thermalized states. The continuous limit of the Galton Board is derived and shown to be a nonlinear Dirac equation (NLDE). The (Galerkin truncated) NLDE evolution is shown to thermalize toward states qualitatively similar to those of the NLOGB. The NLDE conserved quantities are derived and used to construct a stochastic differential equation converging to grand canonical distributions that are shown to reproduce the (micro canonical) NLDE thermalized statistics. Both the NLOGB and the Galerkin-truncated NLDE are thus demonstrated to exhibit spontaneous thermalization.
\end{abstract}

\maketitle

\section{Introduction}

At the fundamental level, quantum theory is linear. Yet, nonlinear models are often useful to take into account interaction in an effective manner. Two examples are the so-called nonlinear optical Galton board (NLOGB) \cite{Perez07} and the nonlinear wave equations describing the dynamics of Bose-Einstein condensates (BEC).
Though the NLOGB is discrete and wave equations are by definition continuous, these models have much in common. Indeed, the NLOGB is essentially a nonlinear quantum walk (QW), and the formal continuous limits of linear QWs are wave equations \cite{knight2003quantum, Dimolfetta2014aa, arrighi2013decoupled, strauch2006relativistic,chandrashekar2010relationship}. Typical such wave equations are the Dirac or the Schr{\"o}dinger equation which nonlinear version, called the Gross-Pitaevskii equation (GPE), is used to model BEC \cite{Berloff2014a}. Also, QW descriptions of BEC have been proposed in \cite{chandrashekar2011disordered,chandrashekar2006implementing}. 

Finally, numerical solutions of continuous wave equations are actually solutions of discrete systems approximating the continuous equations. 

The NLOGB can be seen as a discrete model of nonlinear waves similar to those which propagate in BEC. One can therefore expect the NLOGB to display properties similar to those of the standard nonlinear model of BEC: the GPE. One such property which has until now never been explored on the NLOGB nor, more generally, in the context of QWs and quantum automata, is the so-called spontaneous thermalization. 

In the context of (nonlinear) BEC, microcanonical equilibrium states are well-known to result from long-time integration of the so-called truncated (or Galerkin-projected) Gross-Pitaevskii equation (GPE) and involve a condensation mechanism \cite{krstulovic2011d, krstulovic2011e, shukla2013,Berloff2014a}. 
Furthermore, such thermalization is also known to happen in discretized (rather than spectrally-truncated) GPE \cite{BerloffYoud}.
Classical Galerkin-truncated systems have been studied since the early 50's in fluid mechanics.
In this context, the (time reversible) Euler equation describing spatially-periodic classical ideal fluids is known to admit, when spectrally truncated at wavenumber $k_{\rm max}$, absolute equilibrium solutions with Gaussian
statistics and equipartition of kinetic energy among all Fourier modes 
\cite{lee1952,kraichnan1955,kraichnan1973,orszag1977}.
Furthermore, the dynamics of convergence toward equilibrium involves a direct energy cascade toward small-scales
\cite{Cichowlas2005aa,giorgio09}.

The aim of the present work is to study thermalization phenomena in a spatially-periodic version of the NLOGB and relate it to the thermalization of its (Galerkin-truncated) continuous limit.

The paper is organized as follows. Section \ref{sec:NLDT} is devoted to the definition of the NLOGB model and its numerical solution. The main result of this section is to display and characterize the complex behavior of the log-time regime. Section \ref{sec:NLDE} is devoted to the behavior of the continuous limit (sect. \ref{sec:Cl}), its conserved quantities (sect.\ref{sec:LagC}) and the long-time behavior and thermalization of its Galerkin-truncated version (sect. \ref{sec:LTBandTh}). 
Finally section \ref{sec:Discussion}  is our conclusion. Technical details are given in appendices.


\section{Nonlinear discrete time quantum walk}
\label{sec:NLDT}

\subsection{Fundamentals}
\label{sec:Fund}

Consider a quantum particle endowed with an internal degree of freedom and a lattice on which this particle can move in discrete time. A Discrete Time Quantum Walk (DTQW) is an automaton which conditions the motion of the particle on the state of its internal degree of freedom \cite{Aharonov1993aa}.  Let us remark that whilst a continuous-time version of QW (CTQW) - living in continuous time and discrete space - has been introduced in the literature \cite{kempe2003aa}, we will not deal with it in the present work. 
In this article, we focus on a discrete time nonlinear quantum walks (DTQW) defined on the discrete circle and on particles described by a two components complex wave function. The discrete time spatially periodic quantum walk is defined by the following equations: 
\begin{small}\begin{equation}
\psi^-_{j+1,m} =  
\frac{1}{\sqrt{2}} [ e^{i g \abs{\psi^-_{j,m+1}}^2}\psi^-_{j,m+1}+ e^{i g \abs{\psi^+_{j,m+1}}^2}\psi^+_{j,m+1}]
\label{eq:QW}
\end{equation}\end{small}
\begin{small}
\begin{equation}
\psi^+_{j+1,m} = 
\frac{1}{\sqrt{2}}  [e^{i g \abs{\psi^-_{j,m-1}}^2}\psi^-_{j,m-1} - e^{i g \abs{\psi^+_{j,m-1}}^2}\psi^+_{j,m-1}].
\nonumber
\end{equation}\end{small}
The index $m = 0,...,N-1$  labels points on the discrete circle and the index $j \in \mathbb{N}$ labels instants. At each time $j$ and each point $m$, $\psi^\pm_{j,m}$ are the two components of the wave-function $\Psi_{j,m}$ on a certain space- and time-independent spin basis $(b_-, b_+)$. 
The above finite difference equations were derived in \cite{Perez07}, albeit for QWs on the unrestricted line \footnote{Remark that the unitary evolution operator defining the QW used in \cite{Perez07} is different from the operator used, e.g., in \cite{Dimolfetta2014aa}. The two unitary evolution operators differ in the order in which the quantum coin and translation operator act on the two-component wave function $\Psi_{j,m}$. Note that results on QWs are not modified by interchanging the order on which the quantum coin and the transition operator act on the spinor.}. The parameter $g$ fixes the strength of the nonlinearity. For $g = 0$, equations (\ref{eq:QW}) coincide with the evolution equations of the standard Hadamard walk. 
The particle number at time $j$
\begin{equation}
\Pi_{j}= \sum_m \left(\abs{\psi^-_{j, m}}^2+ \abs{\psi^+_{j, m}}^2 \right) = \sum_m \Pi_{j, m}
\label{eq:PIdisc}
\end{equation}
is independent of $j$ {\sl i.e.} it is conserved by the walk and normalized to 1. We will henceforth denote it by $\Pi$. 
\begin{figure}[h!]
\centering
\includegraphics[width=1.0\columnwidth]{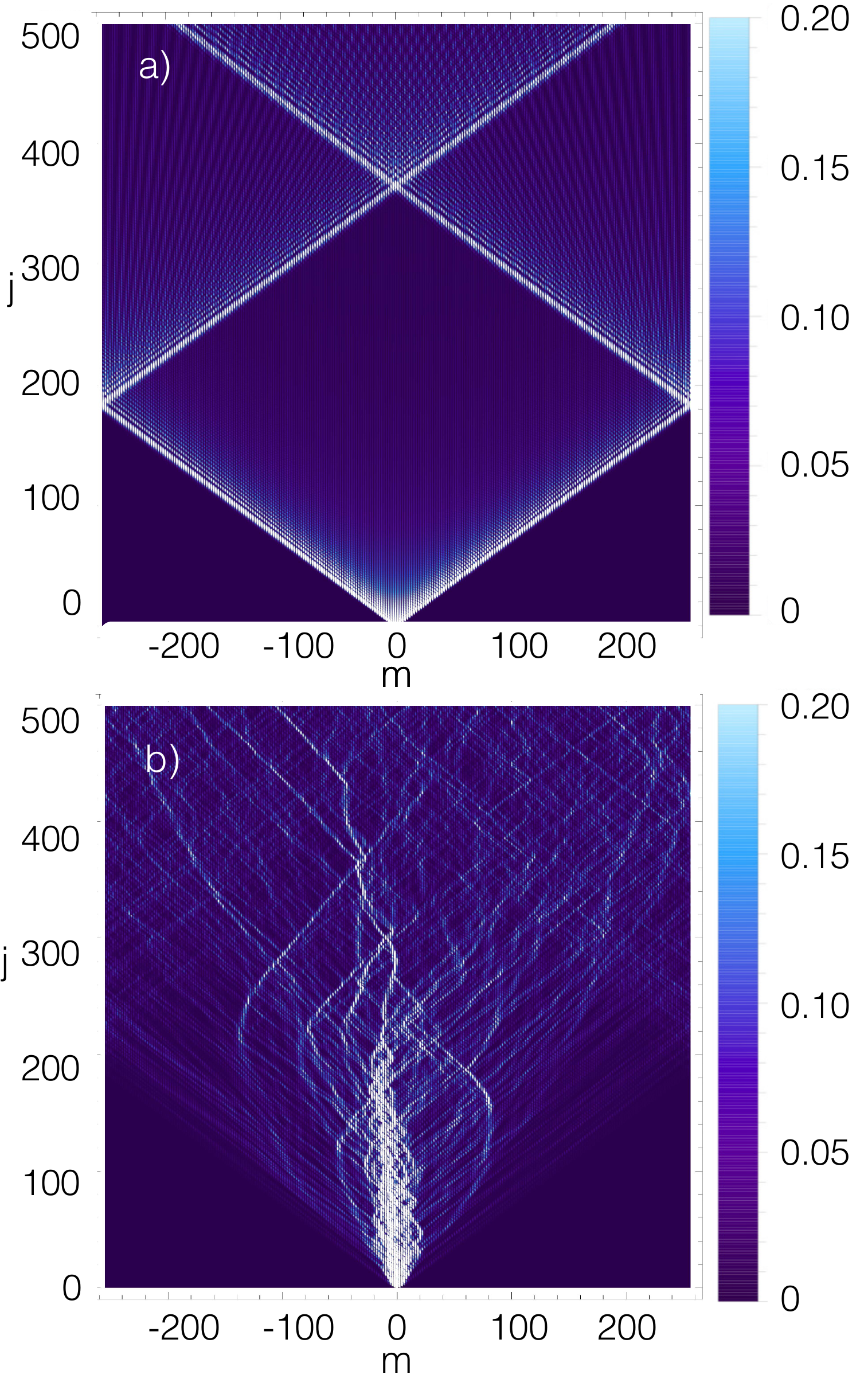}
\caption{(Color online) Density plot of the time evolution of the particle number density $\Pi_{j,m}$, as defined in equation (\ref{eq:PIdisc}), for (a)  the Hadamard DTQW ($g$=0) and the NLOGB  ($g$=10 $\pi$), with a symmetric initial condition $\Psi_{0,m}= \frac{\delta_{0,m}}{\sqrt{2}}(b_-+i b_+)$. Number of grid points N = 512.}
\label{fig:evLin}
\end{figure}


\subsection{Asymptotic behavior of the DTQWs}
\label{secQB}

As displayed in Fig.\ref{fig:evLin}a, the family of DTQWs defined by equations (\ref{eq:QW}) exhibits a very complex dynamics, much richer than the dynamics of the Hadamard walk shown for comparison in Fig.\ref{fig:evLin}b. 

Of particular interest is the $j\to\infty$ asymptotic statistics of the particle number
spatial distribution $\Pi_{j, m}=\abs{\psi^-_{j, m}}^2+ \abs{\psi^+_{j, m}}^2$. Let $\Delta p$ be a positive real number much lower than unity and compute, at all times $j$, the proportion $H_j(p) \Delta p$ of values of the position $m$ for which $\Pi_{jm}$ lies in $(p, p+ \Delta p)$. The quantity $H_j(p)$ is a discrete equivalent of the probability density function (PDF) of  $\Pi_{jm}$ at time $j$.

\begin{figure}[h!]
\includegraphics[width=1.\columnwidth]{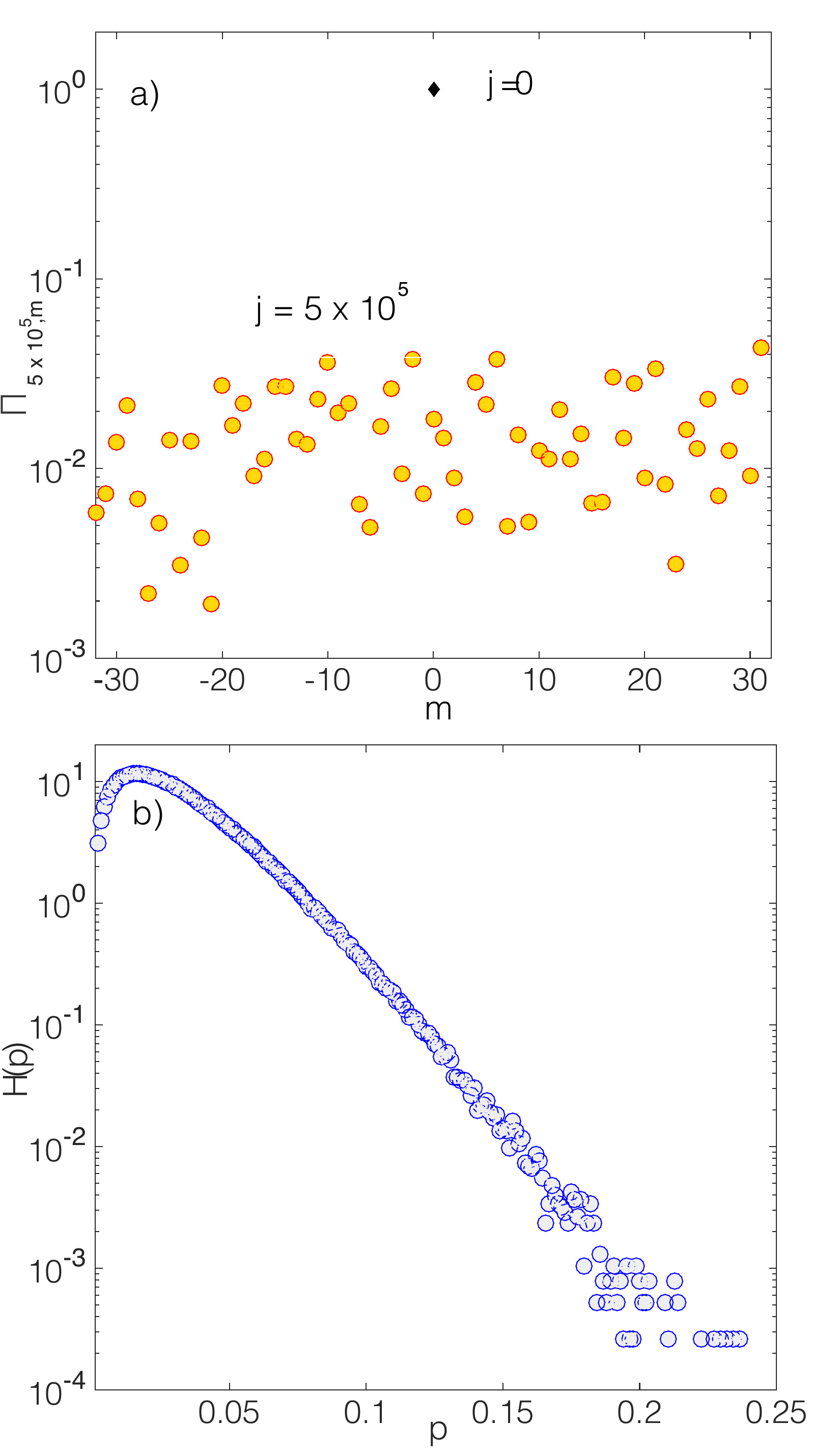}
\caption{(Color online) (a) Log-Lin plot of the particle number density $\Pi_{j,m}$ (see eq.(\ref{eq:PIdisc})) at time $j=5 \times10^5$ (yellow cercles) for the NLOGB with the same initial condition as Fig. \ref{fig:evLin}  (black point) and for $g$=10 $\pi$. Number of grid points $N$ = $64$. (b) Log-Lin PDF $H(p)$ (see text 2nd paragraph in Sec. B) of $\Pi_{j,m}$ at time $j=5 \times10^5$ for the NLOGB with the same initial condition as Fig. \ref{fig:evLin} and $g$=10 $\pi$. Number of grid points $N$ = $64$. 
}
\label{fig:QW2}
\end{figure}

Direct numerical simulations (DNS) (see Fig.\ref{fig:evLin}.a) allow to directly estimate $H_j (p)$. It is found that  $H_j(p)$ tends towards a stationary distribution $H^*(p)$ which depends on the initial condition. Figure Fig. (\ref{fig:QW2}.a) displays how the particle number $\Pi_{j, m}$ typically depends on $m$ at fixed large values of $j$ and Fig. (\ref{fig:QW2}.b) displays $H^*(p)$, estimated by computing histograms as explained above. 

\begin{figure}[h!]
\includegraphics[width=1\columnwidth]{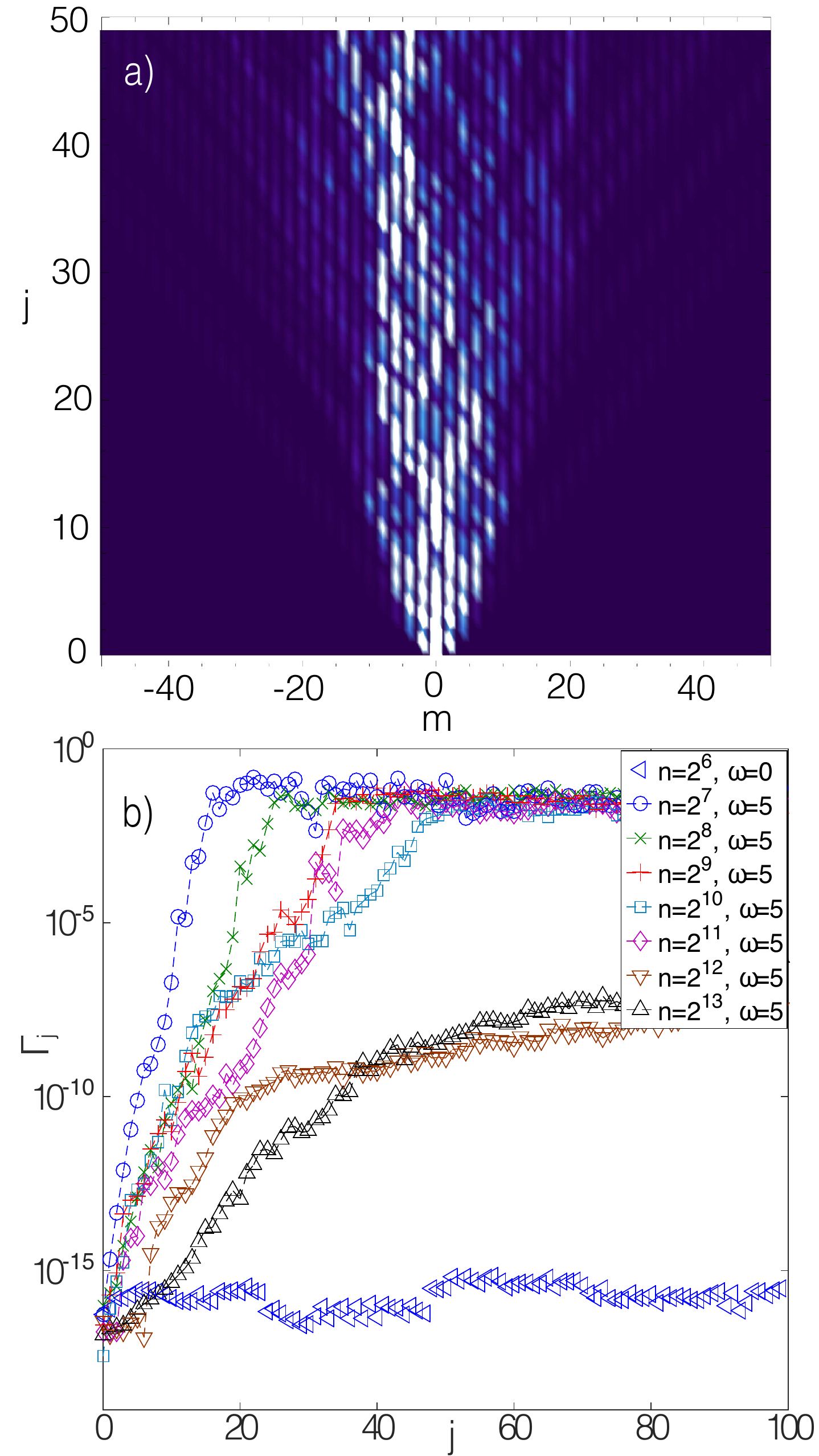}
\caption{(Color online)(a) Density plot of the time evolution of the particle number density $\Pi_{j,m}$  (see eq.(\ref{eq:PIdisc}))  with the same conditions as Fig.\ref{fig:evLin}.b , but in the short time regime. (b) Asymmetry measure $\Gamma_j$ versus time, for different values of the number of grid points. $\Gamma_j$ = $\sum_{m=0}^{N/2-1}\Pi_{j,m} - \sum_{m=N/2}^{N-1}\Pi_{j,m}$ and same conditions as Fig.\ref{fig:evLin}.b.
}
\label{fig:QW3}
\end{figure}

The existence of $H^*(p)$ is typical of nonlinear chaotic systems. These systems also exhibit a great sensitivity towards initial conditions, and this sensitivity is confirmed by DNS of the NLOGB. Indeed, starting a DNS of the Hadamard walk with a symmetric initial condition delivers a numerical solution which is symmetrical at all times, whereas using the same initial condition in a DNS of the NLOGB delivers a numerical solution which is not symmetric (see Fig. \ref{fig:evLin}.a). This symmetry breaking becomes greater with the time $j$ (see Fig. (\ref{fig:QW3}.b) and depends on the resolution of the DNS and the strength of nonlinearities. In particular Fig.\ref{fig:QW3}.b shows that the symmetry breaking starts from the round-off noise \footnote{The Round-off noise or Round-off error is the difference between the computed digital approximation of a number and its exact mathematical value due to rounding.} that is of order $10^{-15}$ in our simulations. We have checked (data not shown) that adding to the initial condition a non-symmetric noise larger than the round-off noise produces the same growth rates for the symmetry breaking, but starting at the higher level of the  added non symmetric noise. This confirms that the symmetry breaking is due entirely to the round-off noise. \\

\section{Nonlinear Dirac equation}
\label{sec:NLDE}
\subsection{
A nonlinear Dirac equation
as continuous limit of the DTQWs}
\label{sec:Cl}
The asymptotic aspects of the NLOGB dynamics can be understood by investigating the continuous limit of these walks. The method employed is the same as in  \cite{Dimolfetta2012aa,Dimolfetta2013aa,Dimolfetta2014aa} and detailed computations are given in Appendix \ref{sec:A1}. The formal continuous limit of the NLQWs read: 
\begin{equation}
\left(\mathbb{I}\partial_T - \sigma_3 \partial_X - \frac{3ig}{4} \mathcal{M}(\Psi, \Psi^\dagger)\right)\Psi  = 0,
\label{eq:CL2}
\end{equation}
with
\begin{equation}
\mathcal{M}(\Psi, \Psi^\dagger) = \Psi^\dagger M \Psi,
\end{equation}
\begin{equation}
M = \mathbb{I} + \frac{\sigma_2}{3}
\end{equation}
where $\mathbb{I}$ is the identity, 
\begin{eqnarray}
\sigma_2 = \left(
\begin{array}{cc}
0 & -i \\
i & 0 \\
\end{array}
\right),
\hspace{0.5cm}
\sigma_3 = \left(
\begin{array}{cc}
1 & 0 \\
0 & -1 \\
\end{array}
\right)
\end{eqnarray}
are the second and third Pauli matrices.
The continuous limit of the NLOGB is thus described by a nonlinear Dirac equation (NLDE). The nonlinearity is confined to the mass term, which depends quadratically on the spinor $\Psi$. Note that (spatially Two-dimensional) NLDE have also been used to describe experimental BEC on 2D hexagonal lattice \cite{haddad2015nonlinear, haddad2011relativistic, haddad2009nonlinear}.

The NLDE (\ref{eq:CL2}) is formally equivalent to Nambu-Jona-Lasinio-like equations (NJLE) (Nambu and Jona-Lasinio, \cite{NJL61,NJL61b,klevansky1992nambu, bijnens1996chiral,zhuang1994thermodynamics}) in 1+1 dimension, which describe a nonlinear interaction between fermions with chiral symmetry. The constant $g$ corresponds to a nonlinear coupling constant and if $g=0$, (\ref{eq:CL2}) degenerates into the Weyl equation.

\begin{figure}[h!]
\centering
\includegraphics[width=1.0\columnwidth]{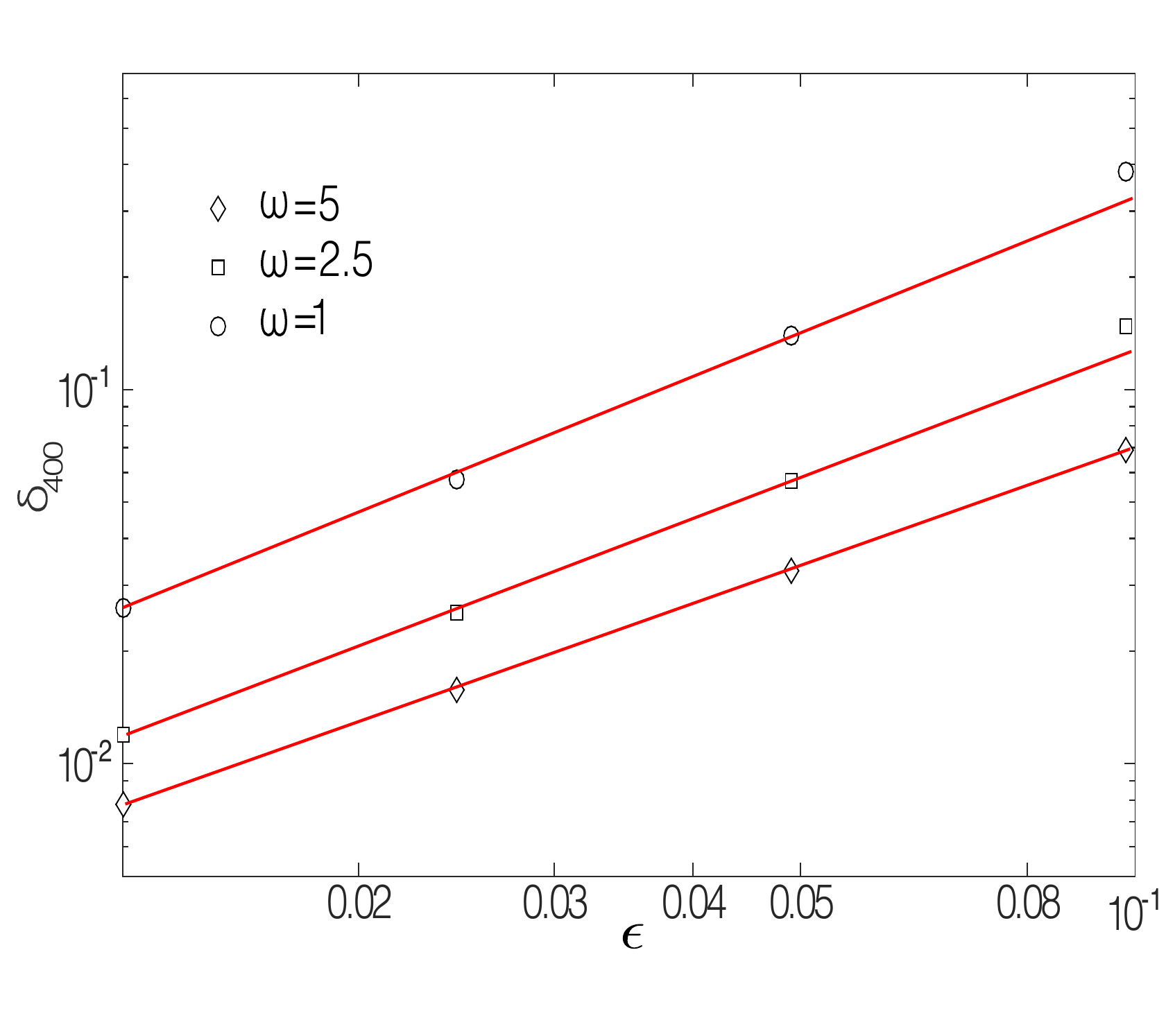}
\caption{(Color online) Log-Log plot of the relative difference $\delta_j$ at time $j=400$, defined as  $\sqrt{\sum^{N-1}_{m=0} \frac{(\Pi_{QW}-\Pi_{D})_{j,m}^2 }{(\Pi_{D})_{j,m}^2}}$, where $\Pi_{QW}$ is the particle number density $\Pi_{j,m}$ obeying to the NLOGB finite difference equations and  $\Pi_{D}$  is the particle number density $\Pi(T_j,X_m)$ obeying to the NLDE. The relative difference is shown for several values of $\epsilon$=$2\pi/N$, (from right to left) N=$2^6$,$2^7$,$2^8$,$2^9$. The initial condition is a symmetric Gaussian distribution $\Psi(0,X_m)= \frac{f(X_m)}{\sqrt{2}}(b_- +i b_+)$, where the Gaussian shape $f(X_m)$ = $\frac{1}{2 \pi \sigma} \exp{\left(- X_m^2/\sqrt{2 \sigma^2}\right)}$ and $\sigma$ =10$\Delta x$.}
\label{fig:diffrel}
\end{figure}

As detailed in Appendix \ref{sec:A2}, the validity of the continuous limit is best confirmed by using Fourier pseudo-spectral methods \cite{orsag00}, which are precise and rather easy to  implement. In particular, Fig. (\ref{fig:diffrel}) displays for different values of $g$ the relative difference between the solution of equations (\ref{eq:QW}) and (\ref{eq:CL2}) as a function of the $\epsilon$ parameter which controls the continuous limit.

\begin{figure}[h]
\centering
\includegraphics[width=1\columnwidth]{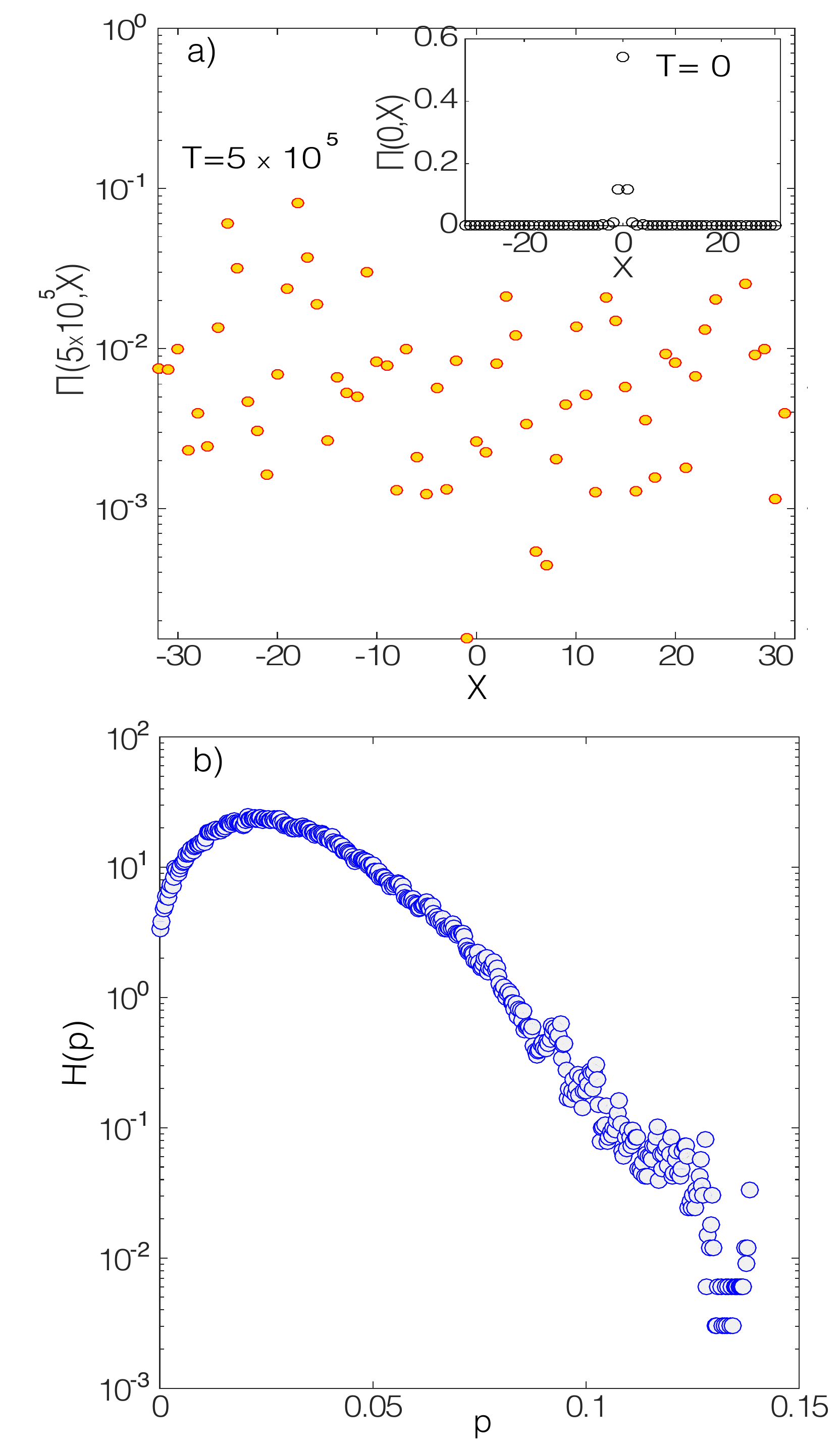}
\caption{(Color online) (a) Log-Lin plot of $\Pi(T,X)$ at time $T=5 \times 10^5$ (red square) obeying the NLDE with the same initial condition of Fig.\ref{fig:diffrel} (black point) for $g$ =10$\pi$. Number of grid points $N$ = $64$. (b) PDF $H(p)$ of $\Pi$. }
\label{fig:densDirac}
\end{figure}

Fig. (\ref{fig:densDirac}) shows the typical profile of the asymptotic $\Pi(T, X)$ and the stationary distribution $H(p)$ of this density, as obtained from a Galerkin-truncated simulation of the NLDE, de-aliased in a way that ensures conservation laws in the truncated system, see Appendix \ref{sec:A2}.
Both plots are strikingly similar to the corresponding plots presented in Fig. \ref{fig:QW2}.a and \ref{fig:QW2}.b obtained by numerically integrating the NLOGB. In other words, the NLOGB and the Galerkin-truncated NLDE seem to have very similar asymptotic behavior. We will now analyze in detail the asymptotic behavior of the Galerkin-truncated NLDE. We will first identify the conserved currents for the NLDE (Section \ref{sec:LagC}) and then show that the asymptotic statistics Galerkin-truncated NLDE is identical to the so-called grand canonical statistics (\ref{sec:LTBandTh}).

\subsection{Lagrangian formulation and conserved quantities}
\label{sec:LagC}

The NLDE derives from the following Lagrangian density:
\begin{equation}
\mathcal{L} (\Psi, \Psi^\dagger)=\frac{i}{2}\left[\overline{\Psi}\gamma^\mu (\partial_\mu \Psi) - (\partial_\mu \overline{\Psi})\gamma^\mu \Psi \right] - \frac{g}{2}
\left(\overline{\Psi}N\Psi\right)^2,
\end{equation}
with
\begin{equation}
N = \gamma_0 + \frac{1}{\sqrt{3}} \gamma_5,
\end{equation}
$\gamma^0 = \sigma_1 = \left(
\begin{array}{cc}
0 & 1 \\
1 & 0 \\
\end{array}
\right)$,  $\gamma^1$= i $\sigma_2$, $\gamma^5$= i $\gamma^0\gamma^1$, ${\overline \Psi} = \Psi^\dagger \gamma^0$ and  $\partial_0=\partial_T$, $\partial_1=\partial_X$. 

There are two conserved currents and these generate three integrals of motion (conserved quantities). The first current is simply the $2$-current  
$J^{\mu}= \overline{\Psi}\gamma^{\mu}\Psi$ associated to the $U(1)$ invariance of the NLDE. The corresponding integral of motion is the total particle number:
\begin{equation}
\Pi[ \Psi, \Psi^\dagger] =\int \Pi(T,X) dX,
\label{eq:PIcont}
\end{equation} 
where $\Pi(T,X) = \Psi^\dagger(T,X) \Psi(T,X)$, which is usually normalized to 1.\\ 
The other current is associated to the space-time translation invariance of the NLDE and is the stress-energy tensor 
\begin{equation}
\mathcal{T}^{\mu \nu} (\Psi, \Psi^\dagger) =\frac{i}{2}[\overline{\Psi}\gamma^\mu(\partial^\nu\Psi) - (\partial^\nu\overline{\Psi})\gamma^\mu\Psi] - \eta^{\mu \nu}\mathcal{L},
\end{equation}
where $\eta^{\mu \nu}$ = $\text{diag}(1,-1)$. The associated conserved quantities are the energy $E$ and the momentum $P$, which are defined by
\begin{equation}
E [ \Psi, \Psi^\dagger] = \int {\mathcal T}^{00}\left( \Psi(T,X), \Psi^\dagger(T,X)\right) dX
\end{equation}
and
\begin{equation}
P  [ \Psi, \Psi^\dagger] = \int {\mathcal T}^{01} \left( \Psi(T,X), \Psi^\dagger(T,X)\right)  dX,
\end{equation}
with
\begin{equation}
\mathcal{T}^{00} (\Psi, \Psi^\dagger) = - \frac{i}{2}\left[\overline{\Psi}\gamma^1 (\partial_X\Psi) - (\partial_X\overline{\Psi})\gamma^1\Psi \right] - \frac{g}{2}
\left(\overline{\Psi}N\Psi\right)^2
\end{equation}
and
\begin{equation}
\mathcal{T}^{01} (\Psi, \Psi^\dagger) =  - \frac{i}{2}  \left[\overline{\Psi}\gamma^0 (\partial_X \Psi) - (\partial_X \overline{\Psi})\gamma^0 \Psi \right].
\end{equation}

\subsection{Thermalization in the Galerkin-truncated NLDE}
\label{sec:LTBandTh}

If one studies the NLDE on the circle, it is natural to write at all times the spinor $\Psi(T,X)$ as a spatial Fourier series and to replace the NLDE by an evolution equation obeyed by the time-dependent Fourier coefficients ${\hat \Psi}(T, k)$. In performing a Galerkin truncation \cite{frisch2008hyperviscosity}, one retains only a {\sl finite} number of these coefficients as dynamical variables, say ${\hat \Psi}(T, k)$ with $k = -\frac{N}{2},...,\frac{N}{2}-1$, and replaces the exact NLDE dynamics by a new dynamics which, at small $k$, approximates at least formally the original NLDE dynamics. By Fourier transforming the ${\hat \Psi}(T, k)$, $k = -\frac{N}{2},...,\frac{N}{2}-1$, back to original physical space ({\sl i.e.} the circle), one obtains a set of N spinors $\Psi_m (T)$, $m = 0..., N-1$, which are to be interpreted as the values $\Psi(T, X_m)$ taken by the spinor field $\Psi(T, X)$ at point $X_m = \frac{2 \pi m}{N}$ (see  Appendix \ref{sec:A2}). The spinors $\Psi(T, X_m)$ are on the same footing as the ${\hat \Psi}(T, k)$, $k = -\frac{N}{2},...,\frac{N}{2}-1$, and can be viewed as the dynamical variables of the Galerkin-truncated NLDE. We now denote by ${\tilde \Psi}(T)$ the collection $\left\{ \Psi_m (T) = \Psi(T, X_m), m = 0, ...N-1\right\}$.

All integrals over space of quantities involving the Dirac field can be replaced by Riemann sums. Thus, the total particle number, the energy and the momentum can now be viewed as functions of the collection $\left({\tilde \Psi}(T), {\tilde \Psi}^*(T)\right)$. These functions will still be denoted by $\Pi$, $P$ and $E$ and are conserves by the Galerkin truncated dynamics, see Appendix \ref{sec:A2}.

\begin{figure}[h!]
\centering
\includegraphics[width=1.\columnwidth]{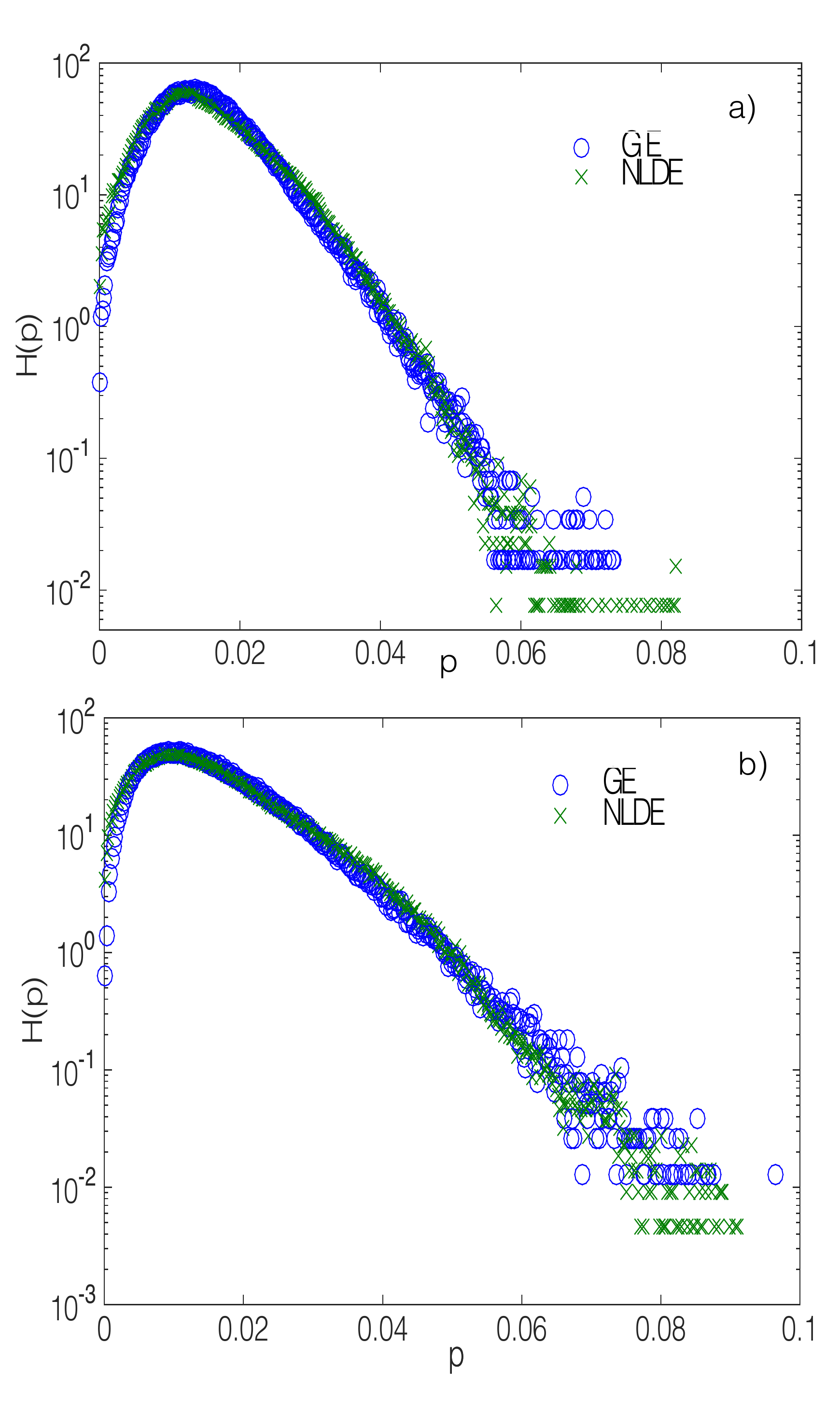}
\caption{(Color online) PDF $H(p)$ of the thermalized state density $\Pi(X)$ for the NLDE and for the stochastic equations (\ref{eq:stoceq}) (GE) for $g= 10 \pi$.  The conserved quantities and the noise coefficient are: a) $E$= -19.8, $P=0$, $\Pi$ = 1 and $\eta$ =1; b) $E$= -17.12, $P=0$, $\Pi$ = 1 and $\eta$ =1.6, at time $T$=312.}
\label{fig:histoD}
\end{figure}

We now introduce two Lagrange multipliers $\mu$ and $\theta$, define 
$H_{\theta \mu} = E - \theta P - \mu \Pi$
and consider the following stochastic differential equations:
\begin{eqnarray}
\frac{d\Psi_m}{dT}  & = & -\ \frac{\partial H_{\theta \mu}}{\partial {\Psi^*_m}}  + \eta \, \xi_m(T) ,\nonumber \\
\frac{d\Psi^*_m}{dT}  & = & -\ \frac{\partial H_{\theta \mu}}{\partial {\Psi_m}}  + \eta \, \xi^*_m(T)
\label{eq:stoceq}
\end{eqnarray}
where $\eta$ is a real coefficient and the $\xi_m(T)$'s are complex independent Gaussian white noises \cite{oksendal2013stochastic} with correlation functions
\begin{equation}
<\xi_m(T) \xi^*_n(T') > =\delta_{m n} \, \delta (T - T'). 
\end{equation}
The density $f$ of this stochastic process obeys the exact Fokker-Planck equation \cite{van2011stochastic,langouche1982functional,krstulovic2011e}
\begin{eqnarray}
\partial_t f = \sum_m \left\{ - \frac{\partial}{\partial \Psi_m} (\frac{\partial H}{\partial \Psi^*_m} f )  - \frac{\partial}{\partial \Psi^*_m} (\frac{\partial H}{\partial \Psi_m} f )+ \right. \nonumber
\\ \left.  \eta^2 \frac{\partial^2 f}{\partial  \Psi_m \Psi_m^*} \right\}.
\label{eq:FP}
\end{eqnarray}
The stationary solution $f_{\eta \theta \mu}$ of this equation is the Gibbs state
\begin{equation}
f_{\eta \theta \mu} 
= \frac{1}{Z_{\eta \theta \mu}}\, \exp\left( - \frac{2}{\eta^2}\, H_{\theta \mu}\right) 
\label{eq:GC}
\end{equation}
 which is the so-called grand canonical distribution with inverse temperature $2/\eta^2$.

We have simulated the stochastic equations (\ref{eq:stoceq}) to numerically obtain 
fields with statistical distributions corresponding to the distribution $f_{\eta \theta \mu}$ and compare it with the asymptotic long-time distribution of the Galerkin-truncated NLDE. 

The comparison between the thermalized fields corresponding to the nonlinear Dirac equation and the Gibbs states corresponding to Eq.(\ref{eq:stoceq})  are displayed in Fig. \ref{fig:histoD} and confirm that the Galerkin-truncated NLDE and the stochastic equations (\ref{eq:stoceq}) are described by very similar distributions (see Appendix \ref{sec:A2}).

Fig. \ref{fig:histoD} has been generated in the following manner. We have first used (\ref{eq:stoceq}) with $\eta=1$ and $\eta=1.6$ to produce two Gibbs states with particle number $\Pi$ fixed to unity and vanishing momentum (see the end of appendix \ref{sec:A2}). The energies of these two states are respectively $E=-19.8$ and $-17.12$. We then have generated initial data for the nonlinear Dirac equation (\ref{eq:CL2}) with the same particle number (equal to unity) and energies. This has been done by multiplying the Gaussian initial field by a suitably chosen spatially dependent phase.


 \section{Discussion}
\label{sec:Discussion}
We have considered the NLOGB confined to the circle and we have shown that the continuous limit of this NLOGB is a NLDE identical to the NJLE-model. Pseudo-spectral numerical simulations reveal that the asymptotic behavior of the NLOGB is similar to the asymptotic behavior of the Galerkin-truncated NLDE and we have shown that the associated asymptotic statistics is identical to the grand-canonical statistics. Thus, both the NLOGB and the Galerkin-truncated NLDE exhibit spontaneous thermalization. Strictly speaking, the NLOGB on an unrestricted line, as presented in \cite{Perez07}, cannot be studied with the same method. In fact the comparison between the NLOGB and the Galerkin-truncated NLDE is possible only when the NLOGB admits a finite number of wave numbers. 

Previous work on other nonlinear quantum walk \cite{shikano2014discrete} suggests that this observed spontaneous asymptotic thermalization is not a particular feature of the systems studied in this article, but will also be encountered in other nonlinear quantum walks, whatever the dimensions of the underlying physical space or of the coin space may be. It is obvious that quantum walks which thermalize will explore space in a very different manner from walks which do not thermalize, and their importance for quantum computing should certainly be explored in depth. In a different direction, it would be interesting to exhibit and analyze spontaneous thermalization in QWs couple to synthetic gauge fields \cite{Dimolfetta2013aa, Dimolfetta2014aa, arrighi2015quantum}.

\appendix

\section{Derivation of continuous limit}\label{sec:A1}

Consider for all $(n,j) \in \mathbb{N}^2$, the collection $W_j^n$ = $(\Psi_{k,m})_{k=nj,m\in\mathbb{Z}}$. This collection represents the state of the NLOGB at 'time' $k=n j$. For any given $n$, the collection $S^n$ = $(W^n_j)_{j\in \mathbb{N}}$ thus represents the entire history of the NLOGB observed through a stroboscope of 'period' $n$. The evolution equations for $S^n$ are those linking $W^n_{j+1}$ to $W_j^n$ for all $j$. The method employed here to obtain the continuous limit of a generic $S^n$ was introduced in \cite{Dimolfetta2013aa,Dimolfetta2014aa}. \\
One first introduces a time-scale $\tau$, a length-scale $\lambda$, an infinitesimal $\epsilon$ and interpret the space-index $m$ as referring to position $x_m = m \epsilon \lambda = m \Delta x$ and the time index $j$ as referring to the instant $t_j = j \epsilon \tau = j \Delta t$. 
The formal continuous limit is obtained expanding the equations defining $S^n$ in Taylor series around $\epsilon = 0$ and by letting 
$\epsilon$ tend to zero. For the limit to exist, all zeroth order terms of the Taylor expansion must identically cancel each other and the differential equation describing the limit is then obtained by equating to zero the non identically vanishing, lowest order contribution.

The original NLOGB $S^1$ does not admit a continuous limit because 
the zeroth order terms do not cancel each other identically. 
The equations defining $S^2$ read:
\begin{scriptsize}
\begin{eqnarray}
\psi^-(t_j+ 2 \Delta t,x_m) =  \frac{1}{2}  [\mathcal{F}[\phi^-(t_j,x_m+\Delta x)] + \mathcal{F}[\phi^+(t_j,x_m-\Delta x)] \nonumber \\ \nonumber
\psi^+(t_j+ 2 \Delta t,x_m) =  \frac{1}{2}  [\mathcal{F}[\phi^-(t_j,x_m-\Delta x)] - \mathcal{F}[\phi^+(t_j,x_m+\Delta x)]
\end{eqnarray}
\end{scriptsize}
where
\begin{footnotesize}
\begin{multline}
\phi^\mp(t_j,x_m)= e^{i g \abs{\psi^-(t_j, x_m+ \Delta x)}^2}\psi^-(t_j, x_m+ \Delta x)\pm\\e^{i g \abs{\psi^-(t_j, x_m+  \Delta x)}^2}\psi^+(t_j, x_m+ \Delta x)
\end{multline}
\end{footnotesize}
and 
\begin{equation}
\mathcal{F}[\phi(t_j,x_m)] = 
e^{i g \abs{\phi(t_j,x_m)}^2} \phi(t_j,x_m).
\end{equation}
These equations admit a formal continuous limit, which reads:
\begin{equation}
( \mathbb{I}\partial_T -  \mathcal{P} \partial_X - \frac{3 i g}{4} \tilde{\mathcal{M}}(\Psi, \Psi^\dagger) )\Psi = 0
\label{eq:CL1}
\end{equation}
where
\begin{equation}
\tilde{\mathcal{M}}(\Psi, \Psi^\dagger) = \Psi^\dagger \tilde{M} \Psi,
\end{equation}
\begin{eqnarray}
\mathcal{P} = \frac{1}{2} 
\left(
\begin{array}{cc}
1 & 1 \\
1 & -1 \\
\end{array}
\right)\hspace{0.5cm} \tilde M = \mathbb{I} - \frac{\sigma_2}{3} 
\end{eqnarray}
and  $T=t/\tau$ and $X=x/\lambda$.

The operator $P$ is self-adjoint and its eigenvalues are $-1$ and $+1$. Two
eigenvectors associated to these eigenvalues are 
\begin{equation}
\mathcal{B}_- = \left(\cos \frac{\theta}{8} \right) b_- + \left(\sin \frac{\theta}{8} \right) b_+
\end{equation}
and
\begin{equation}
\mathcal{B}_+ =  \left(\sin \frac{\theta}{8}\right) b_- - \left( \cos \frac{\theta}{8}\right) b_+.
\end{equation}
The family $(\mathcal{B}_-, \mathcal{B}_+)$ forms an orthonormal basis of the two dimensional spin Hilbert space.
In this new basis, equation (\ref{eq:CL1}) reads:
\begin{equation}
( \mathbb{I}\partial_T -  \sigma_3 \partial_X - \frac{3 i g}{4} \mathcal{M}(\Psi, \Psi^\dagger) )\Psi = 0
\end{equation}
where
\begin{equation}
\mathcal{M}(\Psi, \Psi^\dagger) = \Psi^\dagger M \Psi,
\end{equation}
\begin{eqnarray}
M = \mathbb{I} + \frac{\sigma_2}{3} 
\end{eqnarray}

\section{Numerical Methods}\label{sec:A2}

\begin{figure}[h!]
\centering
\includegraphics[width=0.98\columnwidth]{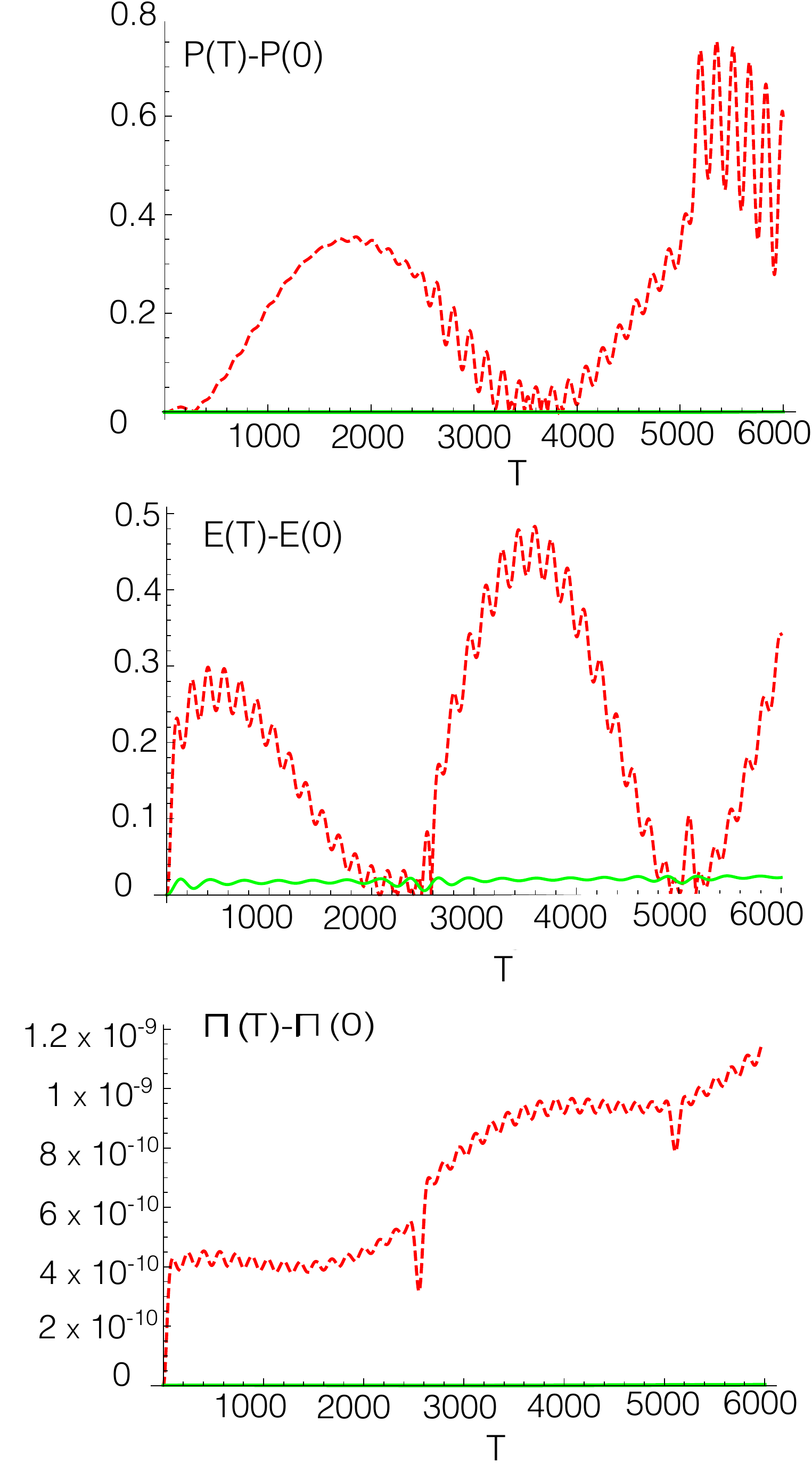}
\caption{(Color online) Time evolution of momentum change (top) $|E(T) - E(T=0)|$, energy change (center) $|P(T) - P(T=0)|$ and particle number change $|\Pi(T)-\Pi(T=0)|$  simulated by a de-aliased pseudo-spectral code for the spatial part and a 4th-order Runge Kutta for the time step. Number of grid points $N$ = $128$. The blue solid lines represents the non de-aliased code.}
\label{fig:Qcons}
\end{figure}

We restrict ourself to $2 \pi$-periodic boundary conditions. A generic field $\Psi(X)$ is thus evaluated on the $N$ collocation points $X_m=2 \pi m/N$, with $m=0,N-1$ as  $\Psi_m=\Psi(X_m)$. The discrete Fourier transforms are standardly defined as $\Psi(X_m) = \sum_{k=-N/2}^{N/2-1} \exp{(i k X_m)}\hat{\psi}_k$ and the inverse $\hat{\psi}_k = \frac{1}{N} \sum_{m=0}^{N-1} \psi(X_m) \exp{(-i k X_m)}$.
These sums can be evaluated in only $N \log(N)$ operations by using Fast Fourier Transforms (FFTs).
Spatial derivatives of fields are evaluated in spectral space by multiplying by $i k$ and products  are evaluated in physical space.
The original QW equations can also be simply cast in this setting, as the translation operator $\Psi_m \to \Psi_{m \pm 1}$ is represented in Fourier space by $\hat{\Psi}_k \to \hat{\Psi}_k \exp{(\pm i k 2 \pi /N)}$. In this setting, the continuous limit is automatically taken when $N$ is increased.  As we can observe in Fig. (\ref{fig:diffrel}) the relative difference scales as expected as $\epsilon$ for different values of $\omega$. \\
However the pseudo-spectral code solving the NLPDEs  generates a problem called aliasing \cite{orsag01}, which means that high $k$-modes alias the amplitudes at lower $k$-modes of the field. In that case the DFT is aliased and in general the fields needs to be de-aliased by proper spectral truncation.
Here, we used the so-called 2/3-rule in all our numerical schemes in the same way as done in reference \cite{krstulovic2011e}. De-aliasing is fundamentally important to preserve the conservation of the Galerkin truncated nonlinear dynamics as we can observe in Fig. (\ref{fig:Qcons}).
Indeed, although it is straightforward to show that Eq.\eqref{eq:CL2} can be written
\begin{eqnarray}
\partial_T \Psi_m = -i \frac{\partial E}{\partial\Psi^*_m}\\
\partial_T \Psi^*_m = i \frac{\partial E}{\partial \Psi_m}
\label{eq:HamD}
\end{eqnarray}
and thus formally conserves the energy, it can be shown that exact conservation requires proper de-aliasing (see appendix of ref. \cite{krstulovic2011e}).

As displayed in Fig.\ref{fig:histoD}, the statistical distributions generated by the NLDE dynamics Eq.\eqref{eq:CL2} and by the stochastic equations (\ref{eq:stoceq}) are really close and this can be justified on very general grounds.

First, by construction, the stochastic equations (\ref{eq:stoceq}) generate the grand canonical distribution (\ref{eq:GC}) that is controlled by the inverse temperature $2/\eta^2$ and the Lagrange multipliers $\mu$ and $\theta$.
On the other hand, as the spectrally-truncated dynamics \eqref{eq:CL2} conserves $\Pi$, $P$ and $E$, its long time behavior should be described by the so-called micro canonical distribution
\begin{equation}
f  \sim  \delta(E-E_{\rm in}) \delta(\Pi-\Pi_{\rm in})  \delta(P-P_{\rm in}).
\end{equation}
that is determined by the values ($E_{\rm in}$,$\Pi_{\rm in}$,$P_{\rm in}$) of the conserved quantities given by the initial condition $\Psi_{\rm in}$. 
As is well-known \cite{landau2013statistical}, under very general circumstances both grand canonical and micro canonical distribution yield similar statistical results (provided that the $2/\eta^2$ and the Lagrange multipliers $\mu$ and $\theta$ have values that correspond to $E_{\rm in}$,$\Pi_{\rm in}$,$P_{\rm in}$). Note that the effect of a fixed value of $\mu$ in Eq. (\ref{eq:stoceq}) amounts, at each time-step, to an overall multiplication of the field by $(1+\mu dt)$. Thus if we want the final result to have a fixed value of the total particle number $\Pi$ this can be obtained setting $\mu$ to zero and, instead, renormalizing the field to the desired value of particle number at each tilme step. 
Fig.\ref{fig:histoD} indicates that, in this case (and zero values for $P$ and $\theta$), both distributions yield identical results for density fluctuations.

\bibliographystyle{unsrt}
\bibliography{NLQW}

\begin{thebibliography}{10}
\bibitem{Note(2)}
The Round-off noise or Round-off error is the difference between the computed
  digital approximation of a number and its exact mathematical value due to
  rounding.
\end{thebibliography}

\end{document}